%% file: main.tex
\documentclass[11pt,letterpaper,conference]{IEEEtran}
\IEEEoverridecommandlockouts
\input{lib/pkg}
\input{lib/cmd}

\begin{document}

\title{Evaluating Smartphone GNSS Accuracy for Geofenced 6 GHz Operations
% Extensive Analysis of GPS Accuracy
% \thanks{\hrule \vspace{4pt} This research is supported in part by the National Science Foundation under grant number CNS-2128489, 2132700, 2220286, 2220292, 2226437, and 2229387.}
}

% IEEE Author block
\author{
\IEEEauthorblockN{
Joshua Roy Palathinkal\IEEEauthorrefmark{1}\IEEEauthorrefmark{3},
Hardani Ismu Nabil\IEEEauthorrefmark{2},
Muhammad Iqbal Rochman\IEEEauthorrefmark{1}, \\
Hossein Nasiri\IEEEauthorrefmark{1},
% S. M. Haider Ali Shuvo\IEEEauthorrefmark{1},
Francis A. Gatsi\IEEEauthorrefmark{1},
and Monisha Ghosh\IEEEauthorrefmark{1}}
\vspace{3pt}
\IEEEauthorblockA{
\IEEEauthorrefmark{1}University of Notre Dame, USA,
\IEEEauthorrefmark{2}Sebelas Maret University, Indonesia.\\ 
\IEEEauthorrefmark{3}Corresponding Author Email: joshuaroy873@gmail.com}
}

% Anonymous author for double-blind review
% \author{\IEEEauthorblockN{Submission \#XXXXX}}

% Copyright notice
% \IEEEpubid{0000--0000/00\$00.00˜\copyright˜2015 IEEE}
% \IEEEoverridecommandlockouts
% \IEEEpubid{\makebox[\columnwidth]{978-1-5386-5541-2/18/\$31.00~\copyright2018 IEEE \hfill} \hspace{\columnsep}\makebox[\columnwidth]{ }}

\maketitle

\begin{abstract}

% The deployment of the 6 GHz spectrum in the U.S. utilizes distinct power categories—such as Standard Power, Low-Power Indoor, and Very Low Power—to protect incumbent users from interference. 
The recently deployed 6 GHz spectrum in the U.S. utilizes distinct power categories, with the latest proposed ``Geofenced Variable Power'' (GVP) category permitting indoor and outdoor operations without continuous Automated Frequency Coordination (AFC) by relying instead on local databases of exclusion zones. Consequently, the safe operation of GVP devices depends entirely on reliable GNSS localization to respect these geofences. However, GNSS accuracy is highly variable and significantly degrades in environments like urban canyons or indoors. This paper presents the first comprehensive empirical study evaluating GNSS reliability specifically for GVP compliance. Utilizing the SigCap Android application, we document and compare GNSS accuracy across an extensive array of real-world conditions, encompassing urban versus suburban landscapes, varying mobility states (stationary, walking, driving), and indoor versus outdoor settings. The results demonstrate that while device hardware causes variations in GNSS accuracy, the operational environment is the primary driver of error. Indoor settings and dense urban areas consistently degrade localization. Moreover, outdoor positions adjacent to buildings often surprisingly produce significant inaccuracies, even near low-elevation structures. We further analyze the contribution of different GNSS constellations to device positioning and show that satellites from non-U.S.-licensed constellations---although currently used in a substantial portion of location fixes---are not permitted for regulatory geolocation under FCC requirements.

\begin{IEEEkeywords}
6 GHz, spectrum sharing, GNSS, accuracy, GVP.
\end{IEEEkeywords}

\end{abstract}

\section{Introduction \& Background} \label{sec:introduction}

\begin{figure}
    \centering
    \includegraphics[width=\linewidth]{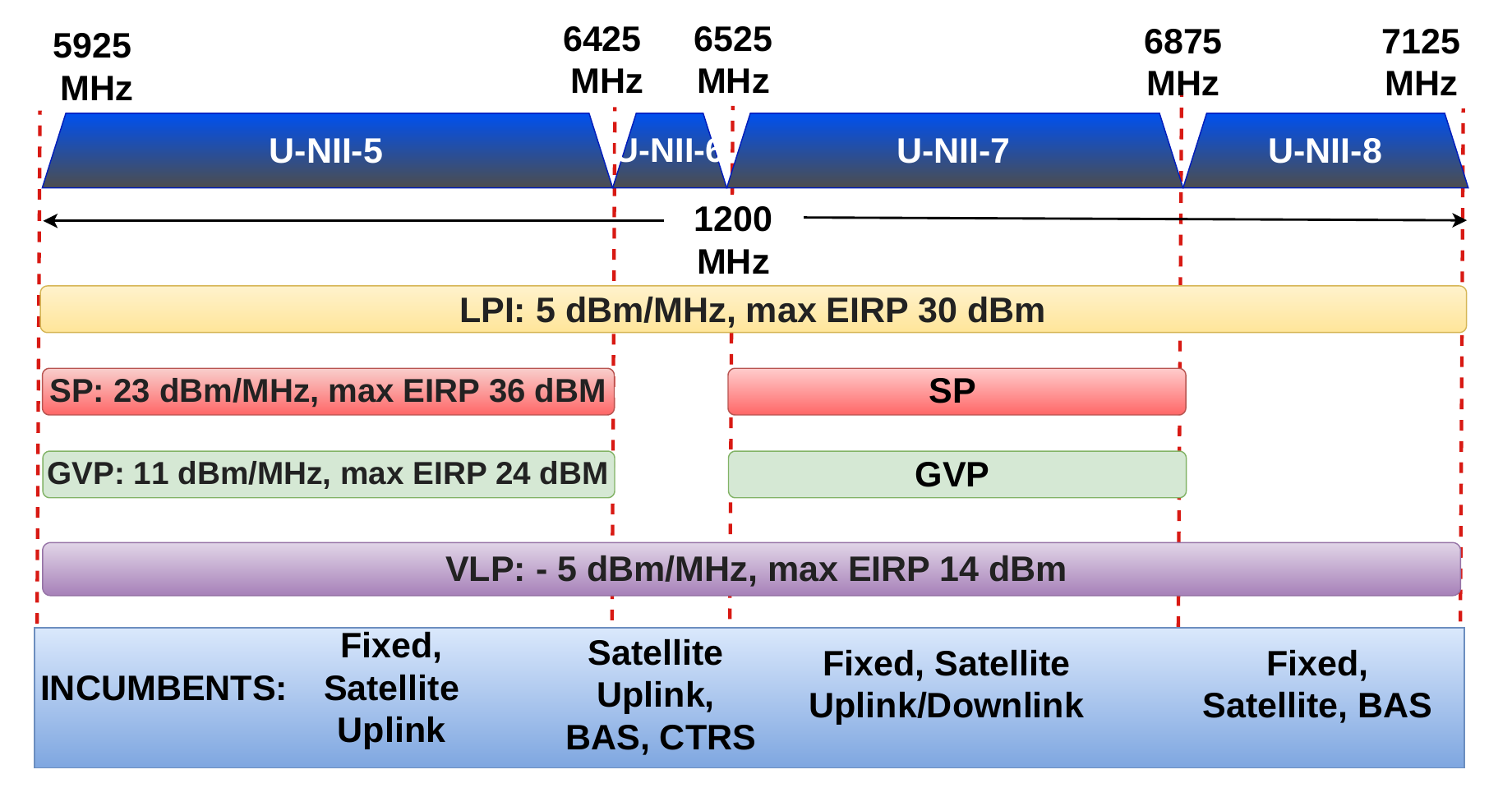}
    \caption{Spectrum Bands and Power Rules in 6 GHz.}
    \label{fig:6ghz_freq_chart}
    % \vspace{-1em}
\end{figure}

The 6~GHz spectrum has been widely deployed in the U.S. under many different power regimes designed to avoid interference with the incumbents (fixed links, satellite, Broadcast Auxiliary Service, etc). These rules establish primary categories: Standard Power (SP), allows the highest power level but requires grants from an Automated Frequency Coordination (AFC) system; Low-Power Indoor (LPI), which mandates lower power and restricts operation to indoor environments without an AFC requirement; and the recently released Very Low Power (VLP) category, enabling access for IoT and wearable devices with the lowest power level and no AFC and indoor requirements. Fig.~\ref{fig:6ghz_freq_chart} summarizes the power and spectrum limitation for each power category.

With a recent Report and Order (R\&O) and Further Notice of Proposed Rulemaking (FNPRM), the Federal Communications Commission (FCC) in the U.S. intends to authorize a new power category: Geofenced Variable Power (GVP)~\cite{fcc_fnprm_gvp}. Devices utilizing this mode will be permitted to operate indoors or outdoors with a Power Spectral Density (PSD) limit of 11 dBm/MHz and a maximum EIRP of 24 dBm. Rather than obtaining authorization from the AFC, these devices will rely on a local database of exclusion zones to avoid interfering with incumbents. Consequently, the ability of GVP devices to operate safely within geofenced zones is heavily dependent on the reliability of the Global Navigation Satellite Systems (GNSS) localization---a technology often mistakenly referred to as GPS. 

% Under FCC certification requirements, the GNSS positioning source must originate only from satellites licensed or approved by the United States~\cite{FCC_TCB_2025}. As a result, signals from non-approved constellations such as GLONASS or BeiDou cannot be used for regulatory compliance or geolocation determination in these systems. This restriction places greater reliance on the performance of the permitted GNSS sources. Because GNSS localization performance varies considerably based on the device capabilities as well as its location (e.g., low accuracy indoors and in urban canyons)~\cite{att_comment_fnprm,zangenehnejad2023gnss}, GVP devices should determine channel availability using real-time reported location accuracy rather than a fixed accuracy value derived from a controlled laboratory environment.

FCC certification guidance further requires that the GNSS source used for geofencing originate only from satellites licensed or approved by the United States under 47~CFR~\S25.137~\cite{FCC_TCB_2025, cfr_25_137}. Signals from non-approved constellations such as GLONASS or BeiDou therefore cannot be used for regulatory geolocation. This constraint increases reliance on the remaining permitted satellites. However, GNSS accuracy varies significantly with device capability and environment (\eg degraded performance indoors or in urban canyons)~\cite{att_comment_fnprm,zangenehnejad2023gnss}. Consequently, channel availability decisions should rely on real-time reported location accuracy rather than fixed accuracy values derived from controlled laboratory environments.

In our prior work, we exploit GNSS inaccuracies---manifested as low received signal strength attributed to building loss and satellite elevation---to predict whether a device is indoors or outdoors~\cite{nasiri2025indoor}. In this work, we extend that investigation to study GVP compliance. While there is a considerably body of work on the empirical analysis of GNSS accuracy in smartphones~\cite{merry2019smartphone,specht2019comparative,fu2020android}, these are focused on a single environment and mobility mode. To the best of our knowledge, this is the first study to comprehensively document and compare GNSS accuracy in real-world environments. We utilize the SigCap Android application across an extensive array of signal environments, \eg, urban \vs suburban, stationary \vs walking \vs driving, and indoor \vs outdoor. Additionally, we analyze the contribution of non-U.S.-licensed satellite constellations to the reported geolocation estimates.

\section{Methodology, Tools and Deployment} \label{sec_deployment}

\begin{table}
\centering
\caption{Data and location summary.}
\label{tab:data_summary}
% \resizebox{\linewidth}{!}{
\begin{tabular}{|C{.25\linewidth}|C{.15\linewidth}|C{.1\linewidth}|C{.1\linewidth}|}
\hline
\textbf{Location} & \textbf{Environment} & \textbf{Devices} & \textbf{\# Datapoints} \\

\hline \hline

\multirow{2}{*}{Chicago, IL} & \multirow{2}{*}{Urban} & S22 & 199\\ \cline{3-4}
& & P8 & 565\\ \hline

\multirow{3}{*}{South Bend, IN} & \multirow{3}{*}{Suburban} & S22 & 2742\\ \cline{3-4}
& & S24 & 1482\\ \cline{3-4}
& & P8 & 215\\ \hline

\multirow{2}{*}{Ames, IA} & Semi- & S22 & 2789\\ \cline{3-4}
& rural & S24 & 1118\\ \hline

San Diego, CA & Urban & S22 & 1740\\ \hline

\multirow{3}{*}{Las Vegas, NV} & \multirow{3}{*}{Urban} & S22 & 1425\\ \cline{3-4}
& & S24 & 1583\\ \cline{3-4}
& & P10 & 869\\ \hline

\end{tabular}
% }
\end{table}

\begin{figure*}
    \begin{subfigure}{.32\textwidth}
        \includegraphics[width=\linewidth]{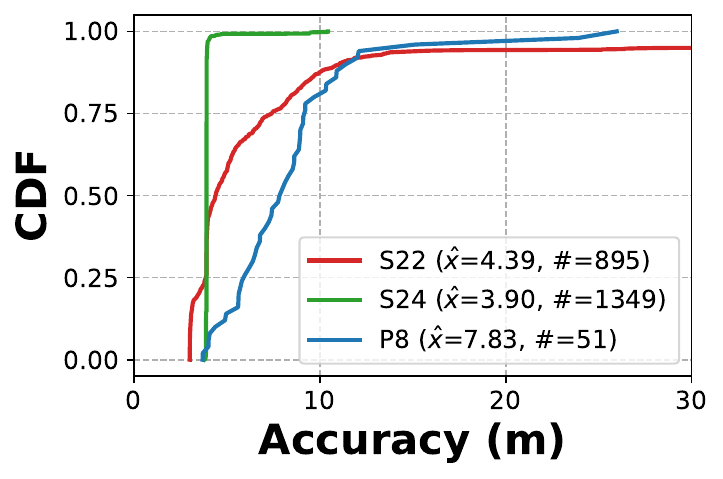}
        \caption{Outdoor walk: Notre Dame Parking}
        \label{figDiffUeNdPark}
    \end{subfigure}
    \hfill
    \begin{subfigure}{.32\textwidth}
        \includegraphics[width=\linewidth]{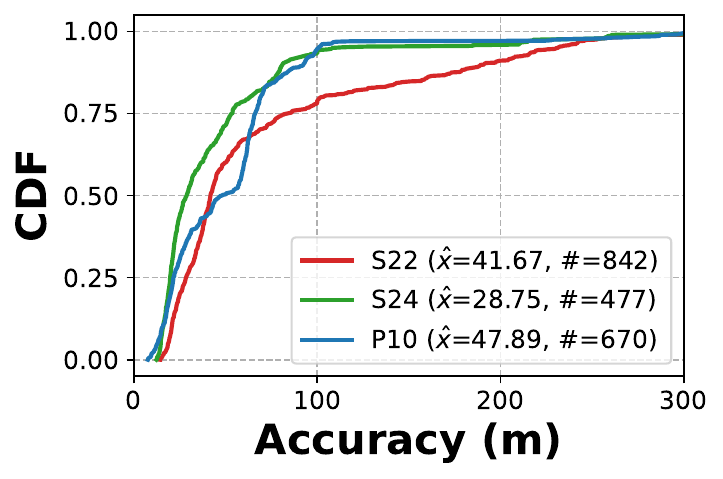}
        \caption{Indoor walk: Las Vegas hotel}
        \label{figDiffUeLasHotelIn}
    \end{subfigure}
    \hfill
    \begin{subfigure}{.32\textwidth}
        \includegraphics[width=\linewidth]{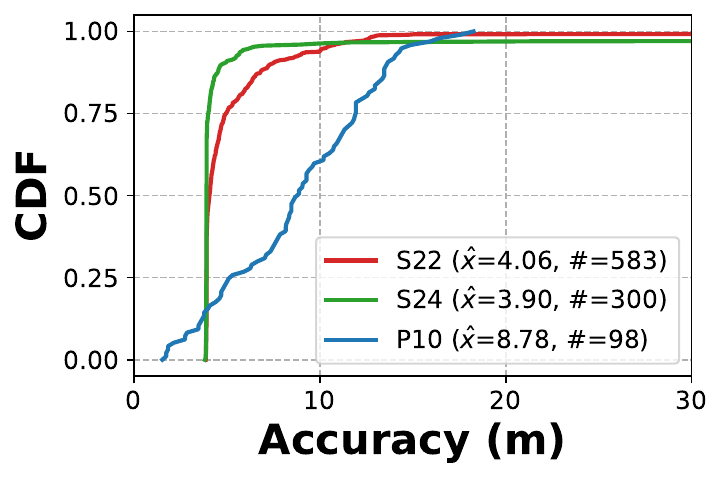}
        \caption{Outdoor drive: Las Vegas city}
        \label{figDiffUeLasDrive}
    \end{subfigure}
    \caption{GNSS accuracy across UE makes and models}
    \label{figDiffUe}
    % \vspace{-1.5em}
\end{figure*}
% \begin{figure*}
%     \centering
%     \subfloat[Outdoor walk: Notre Dame Parking Lot]{
%         \includegraphics[width=.31\textwidth]{figures/resultsv1/figDiffUeNdPark.pdf}
%         \label{figDiffUeNdPark}
%     }
%     \hfill
%     \subfloat[Indoor walk: Las Vegas hotel]{
%         \includegraphics[width=.31\textwidth]{figures/resultsv1/figDiffUeLasHotelIn.pdf}
%         \label{figDiffUeLasHotelIn}
%     }
%     \hfill
%     \subfloat[Outdoor drive: Las Vegas city]{
%         \includegraphics[width=.31\textwidth]{figures/resultsv1/figDiffUeLasDrive.pdf}
%         \label{figDiffUeLasDrive}
%     }
%     \caption{GNSS accuracy across UE makes and models}
%     \label{figDiffUe}
%     % \vspace{-1.5em}
% \end{figure*}

Our empirical evaluation of GNSS accuracy relies on data collected via the SigCap Android application~\cite{dogan2025spectrum}, deployed on several smartphones operating in diverse environments. While SigCap is primarily utilized to crowdsource Wi-Fi and 4G/5G cellular signal information through native Android APIs, its rigorous data-logging pipeline ensures that every measurement is precisely time-stamped and location-tagged. Additionally, each time-stamped measurement includes information about the specific satellites contributing to the positioning solution from the four major GNSS constellations---GPS (United States), Galileo (European Union), BeiDou (China) and GLONASS (Russia). For this study, we extract these associated GNSS coordinates and their corresponding estimated accuracy. Specifically, this accuracy value represents the horizontal radius of a circle centered at the reported coordinates, within which the device's true location lies with a 68\% confidence level~\cite{android_location_SDK}. Moreover, Android utilizes the Fused Location Provider~\cite{android_fused_SDK} which combines signals from GNSS satellites, cellular PCIs, and Wi-Fi BSSIDs: Consequently, the reported accuracy values reflect the most refined localization estimate the device can produce, even in locations where GNSS performance may be suboptimal.

Table~\ref{tab:data_summary} summarizes the measurement locations and smartphone models utilized in this study. Data was collected across three distinct environments: dense urban canyons with high-rise skyscrapers (Chicago, San Diego, and Las Vegas); suburban residential and campus environments (University of Notre Dame, South Bend); and semi-rural areas (Iowa State University, Ames). To ensure hardware diversity, measurements in these locations were taken using four Android models: Google Pixel 8 (P8), Google Pixel 10 Pro (P10), Samsung Galaxy S22+ (S22), and Samsung Galaxy S24+ (S24). In most of the locations, to establish accurate ground truth, each dataset was manually annotated with its corresponding mobility state (\ie, stationary, walking, driving) and environmental context (\ie, indoors, outdoors). To mimic natural usage, device placement varied by mobility state: phones were held at chest level while walking; mounted on the dashboard or placed on the passenger seat while driving; and either held at chest level or resting on a table while stationary. Since measurements were captured at 5-second intervals, the final dataset dataset analyzed in this paper comprises 14,727 data points across all environments and devices, representing approximately 20 hours of data collection.

\section{Results \& Discussions} \label{sec:results}

The results are primarily presented using cumulative distribution function (CDF) curves. For each curve, the legend reports the median value (\(\hat{x}\)) and the corresponding number of data points (\(\#\)).

\subsection{Comparison of Devices} \label{sec:results_ueCompare}

To examine how the phone make and model impacts GPS accuracy, we conducted measurements at three locations with different mobility and environment contexts. The results are summarized in Fig.~\ref{figDiffUe}, where Fig.~\ref{figDiffUeNdPark}, \ref{figDiffUeLasHotelIn} and \ref{figDiffUeLasDrive} correspond to the outdoor walking, indoor walking, and outdoor driving measurements, respectively. Outdoor walking measurements were collected in a parking lot on the University of Notre Dame campus. Indoor walking data were gathered inside a 28\mbox{-}story high-rise hotel-casino in Las Vegas, with an approximate floor area of \(\sim\)6,700~m\(^2\). Outdoor driving measurements were conducted across Las Vegas, with particular emphasis on road segments adjacent to high-rise hotels along the Las Vegas Strip.

The results demonstrate that Samsung UEs exhibit consistent performance between the S22 and S24 models in outdoor environments, with a better median accuracy of under 5~m for both pedestrian and vehicular mobility. In contrast, Google Pixel devices show consistently poorer accuracy than Samsung devices, while remaining broadly similar across Pixel models (P8 vs. P10). We note that not all Pixel models were available across all outdoor scenarios; in particular, both P8 and P10 were not measured in the outdoor walking dataset (Fig.~\ref{figDiffUe}) and the outdoor driving dataset (Fig.~\ref{figDiffUeLasDrive}). For indoor walking (Fig.~\ref{figDiffUeLasHotelIn}), GPS accuracy degrades substantially across all devices. Under these conditions, the Samsung models diverge more noticeably (41.67~m for S22 vs. 28.75~m for S24), and the P10 yields the largest error (47.89~m).
These results highlight the variability of localization accuracy across different devices, a factor that must be carefully considered in the practical implementation of GVP rules.

\subsection{Comparison of Mobility Modes} \label{sec:results_mobilityCompare}

\begin{figure}
    \centering
    \includegraphics[width=0.64\linewidth]{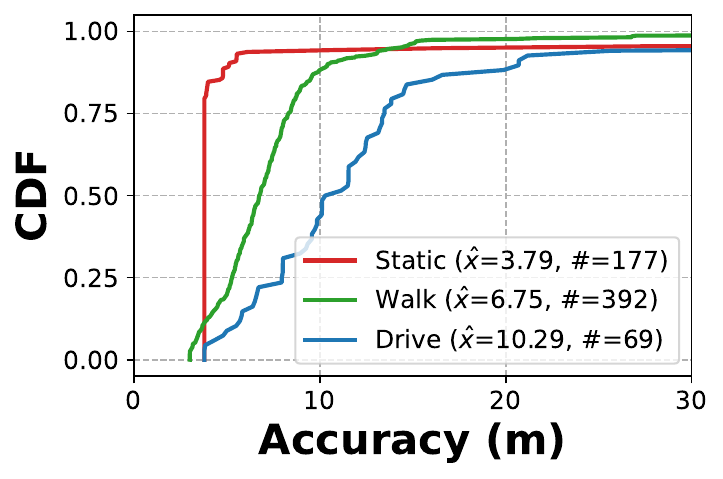}
    \caption{Outdoor mobility: San Diego downtown}
    \label{figDiffMobility}
\end{figure}

For \S\ref{sec:results_mobilityCompare} and \S\ref{sec:results_envCompare}, we use the Samsung S22 as the representative UE unless stated otherwise. Results for different outdoor mobility modes are shown in Fig.~\ref{figDiffMobility}. The measurements were collected in San Diego: static data were obtained at a hotel parking lot just outside the downtown region; walking data were collected along a roadway adjacent to the San Diego Bay; and driving data were collected while traversing downtown streets characterized by high-rise buildings. The results show clear variation in GPS accuracy across mobility modes, with the best performance under static conditions (median 3.79~m), followed by walking (6.75~m), and the poorest performance under driving (10.29~m).

\subsection{Comparison of Environments}  \label{sec:results_envCompare}

\subsubsection{Building Effect}

\begin{figure}
    \centering
    \begin{subfigure}{\linewidth}
        \centering
        \includegraphics[width=0.4\linewidth]{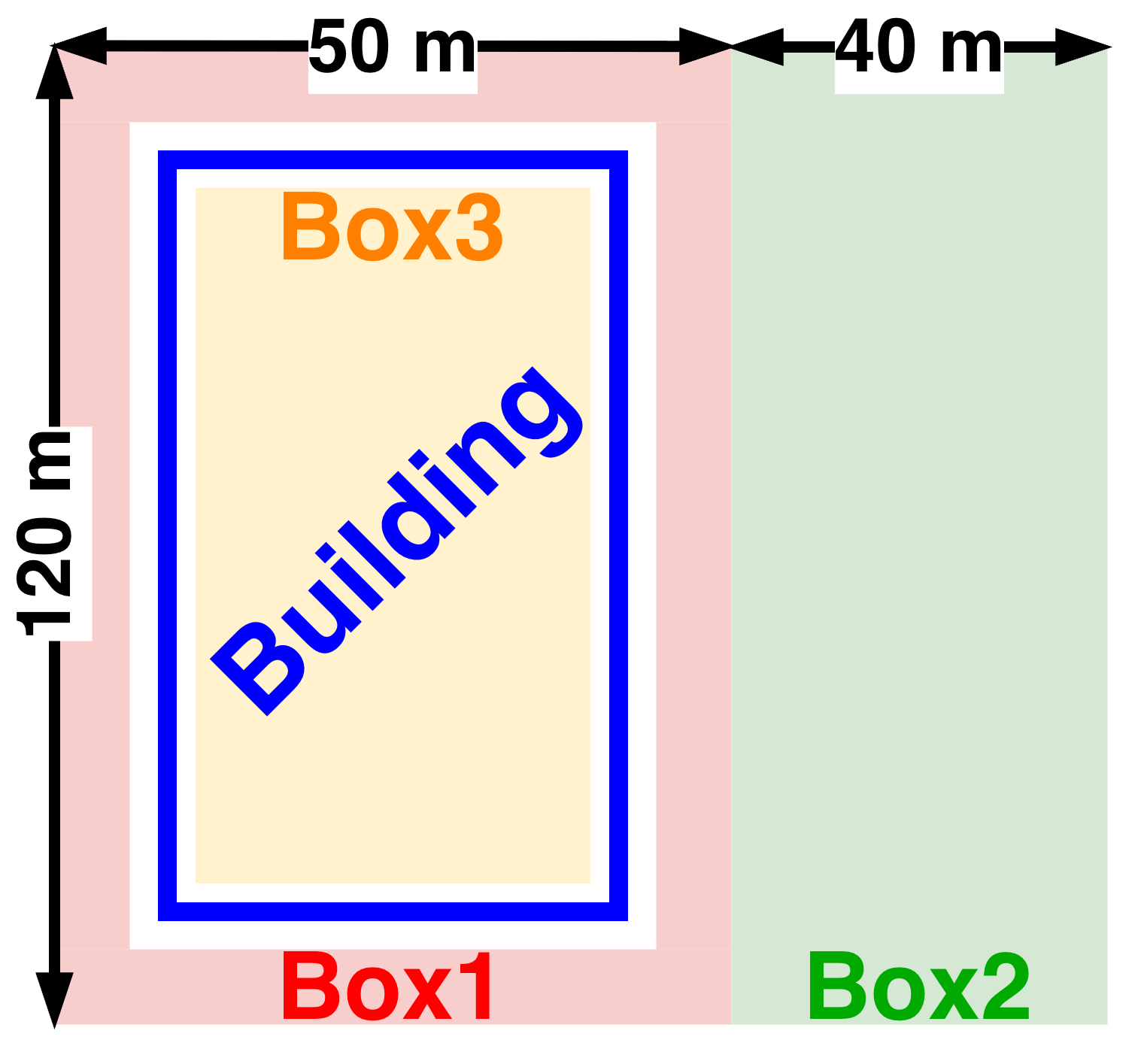}
        \caption{Regions overview}
        \label{figNdDebartDesc}
    \end{subfigure}
    \begin{subfigure}{\linewidth}
        \centering
        \includegraphics[width=0.64\linewidth]{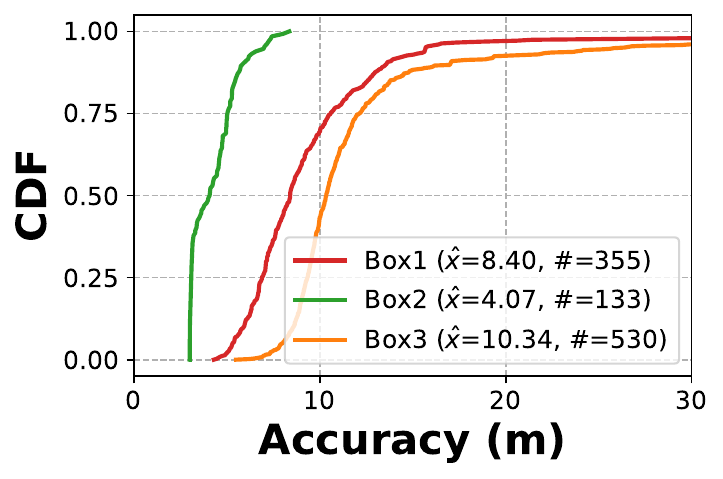}
        \caption{Walk: Notre Dame}
        \label{figNdDebart}
    \end{subfigure}
    \caption{GPS accuracy by environment (Notre Dame)}
    \label{figDiffEnv1}
\end{figure}

Fig.~\ref{figDiffEnv1} summarizes GPS accuracy during walking measurements collected in multiple regions around a building on the University of Notre Dame campus. As illustrated in Fig.~\ref{figNdDebartDesc}, we consider three regions: \textit{Box1}, a path immediately adjacent to the building (with the building wall on one side and open space on the other); \textit{Box2}, a path approximately 15~m away from the building; and \textit{Box3}, an indoor path on the first floor. The building is characterized by thick walls. As shown in Fig.~\ref{figNdDebart}, moving from \textit{Box1} to \textit{Box2} yields a substantial improvement in median accuracy, from 8.4~m to 4.07~m. This indicates that proximity to the building degrades GPS accuracy by approximately 4.33~m. In \textit{Box3} (indoors), the median accuracy further degrades to 10.34~m, corresponding to a indoor-to-outdoor median accuracy reduction of approximately 6.27~m relative to the outdoor \textit{Box2} region.

\subsubsection{Urban and Semi-Rural Environments}

A similar indoor-to-outdoor transition effect is also evident in the Las Vegas measurements for the S22. Comparing indoor walking in Fig.~\ref{figDiffUeLasHotelIn} to outdoor driving in Fig.~\ref{figDiffUeLasDrive}, the median GPS error decreases by 37.61~m. The larger magnitude of this transition relative to the Notre Dame results is plausibly attributable to differences in the built environment, as the Las Vegas hotel structure is substantially taller (28 floors) than the Notre Dame campus building (3 floors).

% \begin{figure}
%     \centering

%     \subfloat[Regions overview]{
%         \centering
%         \includegraphics[width=0.4\linewidth]{figures/resultsv1/figNdDebartDesc.png}
%         \label{figNdDebartDesc}
%     }

%     \subfloat[Walk: Notre Dame]{
%         \centering
%         \includegraphics[width=0.64\linewidth]{figures/resultsv1/figNdDebart.pdf}
%         \label{figNdDebart}
%     }

%     \caption{GPS accuracy by environment (Notre Dame)}
%     \label{figDiffEnv1}

% \end{figure}

\begin{figure}
    \centering
    \begin{subfigure}{\linewidth}
        \centering
        \includegraphics[width=0.64\linewidth]{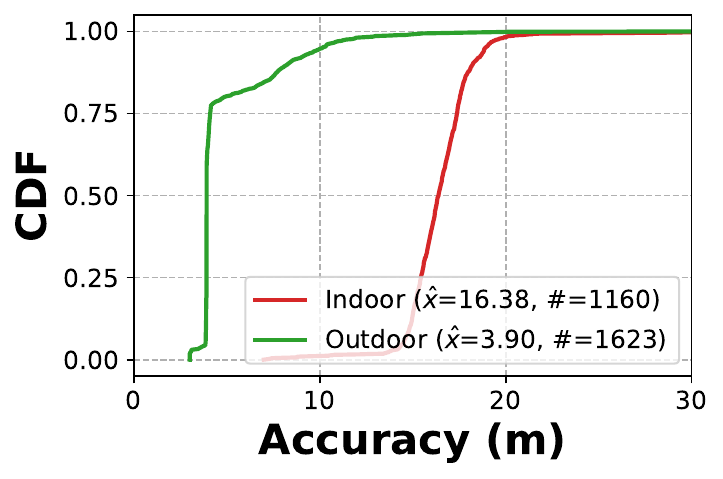}
        \caption{Walking regions: Iowa State University}
        \label{figIowa}
    \end{subfigure}
    \begin{subfigure}{\linewidth}
        \centering
        \includegraphics[width=0.64\linewidth]{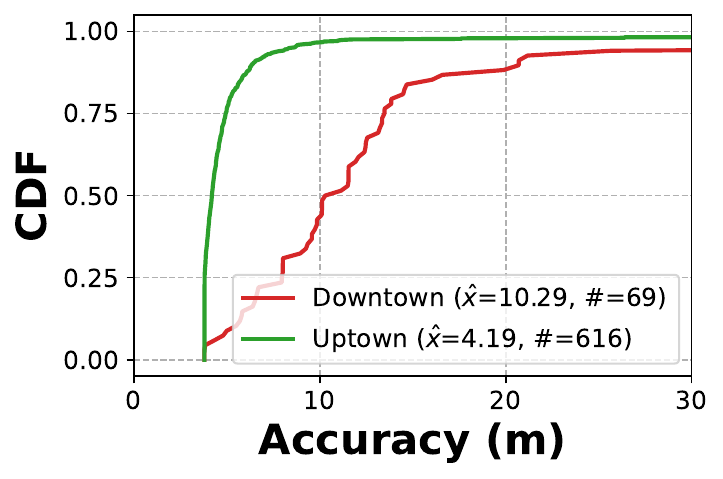}
        \caption{Drive: San Diego}
        \label{figSanDiegoDrive}
    \end{subfigure}
    \caption{GPS accuracy by environment (Iowa \& San Diego)}
    \label{figDiffEnv2}
\end{figure}

Another two examples comparing semi-rural and urban environments are shown in Fig.~\ref{figDiffEnv2}. Fig.~\ref{figIowa} presents measurements collected at Iowa State University, which can be characterized as a semi-rural setting with relatively low building heights and sparse building density. The indoor measurements were taken inside a five-story building with extensive glass facades, which can introduce additional attenuation and multi-path effects, while the outdoor measurements correspond to walking routes around campus. As shown in Fig.~\ref{figIowa}, the median GPS error improves substantially in the indoor-to-outdoor transition, decreasing from 16.38~m indoors to 3.9~m outdoors---an improvement of 12.48~m.

% \begin{figure}
%     \centering

%     \subfloat[Walking regions: Iowa State University]{
%         \centering
%         \includegraphics[width=0.64\linewidth]{figures/resultsv1/figIowa.pdf}
%         \label{figIowa}
%     }

%     \subfloat[Drive: San Diego]{
%         \centering
%         \includegraphics[width=0.64\linewidth]{figures/resultsv1/figSanDiegoDrive.pdf}
%         \label{figSanDiegoDrive}
%     }

%     \caption{GPS accuracy by environment (Iowa \& San Diego)}
%     \label{figDiffEnv2}

% \end{figure}

Fig.~\ref{figSanDiegoDrive} reports driving measurements from two regions in San Diego: downtown and uptown. Downtown represents an urban environment with dense high-rise structures, whereas uptown is more suburban, dominated by residential buildings, with the driving route primarily along Interstate~5. Consistent with the expected impact of dense urban infrastructure on GPS accuracy, downtown exhibits poorer median accuracy (10.29~m) relative to uptown (4.19~m), corresponding to a median difference of 6.1~m.

\subsection{Constellation Selection \& Regulatory Requirements} 
\label{sec:sat_analysis}

In addition to environmental factors, geofencing operations in the 6~GHz band are subject to constellation restrictions. As discussed in \S\ref{sec:introduction}, FCC certification guidance restricts geofencing localization to satellites licensed or approved by the United States. Under these rules, constellations such as GLONASS and BeiDou are not permitted for geofencing validation~\cite{FCC_TCB_2025}. 

All satellite samples analyzed in this section correspond to measurements where the \texttt{usedInFix} parameter is set to \texttt{True}, indicating that the satellite contributed to the device’s geoposition calculation. Our empirical data indicate that these restrictions exclude a substantial portion of available signals. 

\begin{figure}
    \centering
    \begin{subfigure}{\linewidth}
        \centering
        \includegraphics[width=0.7\linewidth]{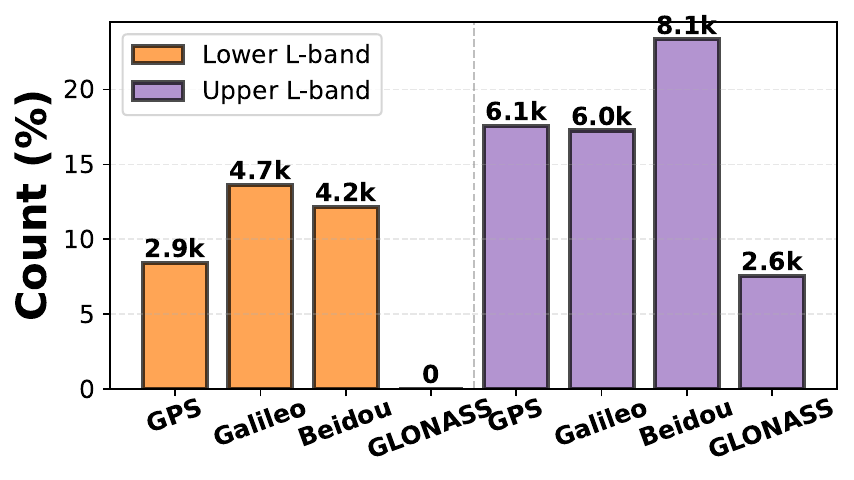}
        \caption{Histogram of satellite samples across L-bands}
        \label{figSatelliteCountHist}
    \end{subfigure}
    \begin{subfigure}{\linewidth}
        \centering
        \includegraphics[width=0.64\linewidth]{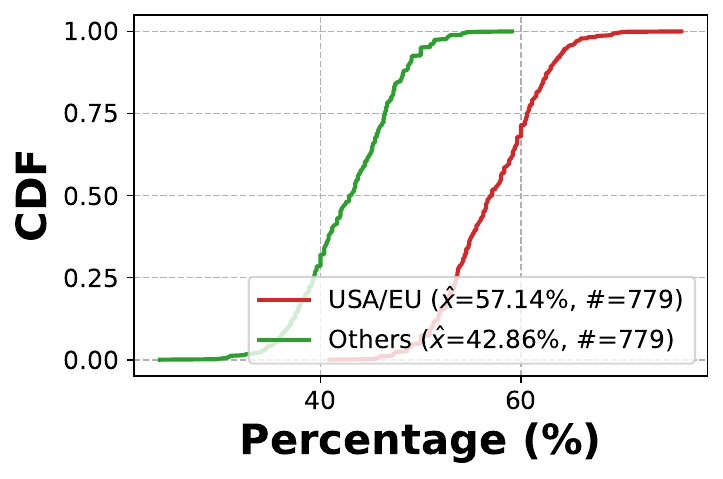}
        \caption{CDF of constellation contribution in location fixes}
        \label{figSatelliteCountCdf}
    \end{subfigure}
    \caption{Histogram and CDF of satellite counts across L-bands and constellations}
    \label{figSatelliteCount}
\end{figure}

Fig.~\ref{figSatelliteCountHist} shows the distribution of satellite samples across L-bands for all observations in the dataset. Because multiple satellite measurements are recorded at each timestamp, this figure reflects the total number of satellite observations rather than the number of distinct location fixes. As shown, BeiDou satellites are the most frequently observed signals in the upper L-band, with 8.1\(\times10^3\) samples. In the lower L-band, BeiDou ranks second with 4.2\(\times10^3\) samples, closely following Galileo, which has the highest count at 4.7\(\times10^3\) samples.

% \begin{figure}
%     \centering

%     \subfloat[Histogram of satellite samples across L-bands]{
%         \centering
%         \includegraphics[width=0.7\linewidth]{figures/resultsv1/figSatelliteCountHist.pdf}
%         \label{figSatelliteCountHist}
%     }

%     \subfloat[CDF of constellation contribution in location fixes]{
%         \centering
%         \includegraphics[width=0.64\linewidth]{figures/resultsv1/figSatelliteCountCdf.pdf}
%         \label{figSatelliteCountCdf}
%     }

%     \caption{Histogram and CDF of satellite counts across L-bands and constellations}
%     \label{figSatelliteCount}
% \end{figure}

To better capture the constellation composition of actual location fixes, Fig.~\ref{figSatelliteCountCdf} aggregates samples at the timestamp level. Specifically, multiple satellite observations recorded at the same timestamp are reduced to a single positioning instance, and the percentage contribution of different constellations is computed for that fix. In this figure, \textit{USA/EU} refers to satellites from the GPS (United States) and Galileo (European Union) constellations, which are permitted for regulatory geolocation, while \textit{Others} refers to satellites from the BeiDou (China) and GLONASS (Russia) constellations. The results show that BeiDou and GLONASS together account for a median of 42.86\% of the satellites used in location fixes.

The FCC restriction imply that, for geofencing purposes, devices must disregard approximately 43\% of the satellites currently used in location fixes. Consequently, localization may need to rely solely on the permitted subset of satellites, which may affect the accuracy of the coordinates used for geofence validation.

\section{Conclusions \& Future Research} \label{sec:conclusions}

The addition of GVP devices to the 6~GHz ecosystem will enable new applications and use-cases. However, since the 6~GHz band is shared with incumbents such as fixed microwave links, it is imperative that sharing methodologies be robust while enhancing spectrum utilization. Localization techniques rely on GNSS, cellular, and Wi-Fi signals to enhance positioning accuracy, and because the availability of these signals depends on the local environment, it is essential that GVP rules require channel selection to consider the device's measured localization accuracy rather than a static, lab-certified value that does not reflect real-world conditions. In addition, regulatory restrictions on GNSS constellations may further reduce the number of satellites available for positioning, which can affect the accuracy used for geofence validation.

This paper presents a first-of-its-kind analysis of extensive real-world measurements demonstrating the variability of location accuracy across diverse environments, and examines the role of different GNSS constellations in device positioning. These results support the need for accuracy-aware channel selection in GVP devices. As GVP sharing mechanisms mature, future work will explore methods to improve spectrum sharing while ensuring reliable incumbent protection.

% \section*{\centering{Acknowledgements}}

% This work is supported by NSF grants CNS-XXXXXXX and AST-XXXXXXX.

\bibliographystyle{IEEEtran}
\bibliography{main}

\end{document}

%% file: lib/pkg.tex
\usepackage{amsmath,amssymb,amsfonts}
\usepackage{gensymb}
\usepackage{array}
\usepackage{algorithmic}
\usepackage{graphicx}
\usepackage{textcomp}
\usepackage{xcolor}
\usepackage{xspace}
\usepackage{enumitem}
\usepackage{url}
\usepackage{breakurl}
\usepackage{caption}
\usepackage{subcaption}
\usepackage{multirow}
\usepackage[normalem]{ulem}
\usepackage[numbers,sort&compress]{natbib}
\usepackage[linesnumbered,ruled,lined,boxed]{algorithm2e}

%% file: lib/cmd.tex
\graphicspath{{figures/}}
\usepackage[left=0.65in,right=0.65in,top=0.75in,bottom=1.1in]{geometry}
\SetKwInput{KwInput}{Input}                % Set the Input
\SetKwInput{KwOutput}{Output}              % set the Output

\newcolumntype{L}[1]{>{\raggedright\let\newline\\\arraybackslash\hspace{0pt}}m{#1}}
\newcolumntype{C}[1]{>{\centering\let\newline\\\arraybackslash\hspace{0pt}}m{#1}}
\newcolumntype{R}[1]{>{\raggedleft\let\newline\\\arraybackslash\hspace{0pt}}m{#1}}

\newcommand{\ie}{\textit{i.e.,}\xspace}
\newcommand{\eg}{\textit{e.g.,}\xspace}

\newcommand{\vs}{\textit{vs.}\xspace}

%% file: main.bib
@inproceedings{nasiri2025indoor,
  title={Indoor/Outdoor Spectrum Sharing Enabled by {GNSS}-based Classifiers},
  author={Nasiri, Hossein and Rochman, Muhammad Iqbal and Ghosh, Monisha},
  booktitle={MILCOM 2025-2025 IEEE Military Communications Conference (MILCOM)},
  pages={330--337},
  year={2025},
  organization={IEEE}
}

@misc{fcc_fnprm_gvp,
  author = {{Federal Communications Commission}},
  title = {{Fourth Report \& Order and Third Further Notice of Proposed Rulemaking; Expanding Unlicensed Operations in the 6 {GHz} Band}},
  year = {2026},
  month = {Jan.},
  howpublished={Retrieved from \url{https://docs.fcc.gov/public/attachments/DOC-417577A1.pdf}}
}

@misc{att_comment_fnprm,
  title = {{Comments Of AT\&T Services, Inc.; Expanding Flexible Use in Mid-Band Spectrum Between 3.7 and 24 {GHz}}},
  year = {2024},
  month = {Mar.},
  howpublished={Retrieved from \url{https://www.fcc.gov/ecfs/document/10327028492354/1}}
}

@misc{android_location_SDK,
  author = {{Android Developers}},
  title = {{Location - API Reference}},
  year = {2026},
  month = {Feb.},
  note={Accessed: Feb. 2026},
  howpublished={Retrieved from \url{https://developer.android.com/reference/android/location/Location}}
}

@article{dogan2025spectrum,
  title={Spectrum sharing characterization using smartphones: Exploring 6 {GHz} sharing through large-scale {Wi-Fi 6E} measurements},
  author={Do{\u{g}}an-Tusha, Seda and Tusha, Armed and Rochman, Muhammad Iqbal and Nasiri, Hossein and Ghosh, Monisha},
  journal={IEEE Communications Magazine},
  volume={63},
  number={2},
  year={2025},
  publisher={IEEE}
}

@misc{android_fused_SDK,
  author = {{{Android Developers}}},
  title = {{Fused Location Provider API}},
  year = {2026},
  month = {Feb.},
  note={Accessed: Feb. 2026},
  howpublished={Retrieved from \url{https://developers.google.com/location-context/fused-location-provider}}
}

@inproceedings{fu2020android,
  title={{Android raw GNSS measurement datasets for precise positioning}},
  author={Fu, Guoyu Michael and Khider, Mohammed and Van Diggelen, Frank},
  booktitle={Proceedings of the 33rd international technical meeting of the satellite division of the Institute of Navigation (ION GNSS+ 2020)},
  pages={1925--1937},
  year={2020}
}

@article{specht2019comparative,
  title={{Comparative analysis of positioning accuracy of GNSS receivers of Samsung Galaxy smartphones in marine dynamic measurements}},
  author={Specht, C and Dabrowski, PS and Pawelski, J and Specht, M and Szot, T},
  journal={Advances in Space Research},
  volume={63},
  number={9},
  pages={3018--3028},
  year={2019},
  publisher={Elsevier}
}

@article{merry2019smartphone,
  title={{Smartphone GPS accuracy study in an urban environment}},
  author={Merry, Krista and Bettinger, Pete},
  journal={PloS one},
  volume={14},
  number={7},
  pages={e0219890},
  year={2019},
  publisher={Public Library of Science}
}

@misc{zangenehnejad2023gnss,
  title={{GNSS observation generation from smartphone android location API: performance of existing apps. Issues Improv Sens 23 (2): 777}},
  author={Zangenehnejad, F and Jiang, Y and Gao, Y},
  year={2023}
}

@misc{FCC_TCB_2025,
  author       = {{{Federal Communications Commission}}},
  title        = {Administrative Notes and Publication Update -- TCB Workshop},
  howpublished = {\url{https://www.fcc.gov/sites/default/files/13-Administrative-Notes-and-Publication-Update_Apr_2025_Final-TCB.pdf}},
  month        = apr,
  year         = {2025},
  note         = {{Office of Engineering and Technology, Laboratory Division}}
}

@misc{cfr_25_137,
  author = {{Federal Communications Commission}},
  title = {{47 C.F.R. § 25.137 -- Requests for U.S. Market Access Through Non-U.S.-Licensed Space Stations}},
  howpublished = {\url{https://www.ecfr.gov/current/title-47/chapter-I/subchapter-B/part-25/subpart-B/subject-group-ECFR34f9987bbfcd1a8/section-25.137}},
  note = {Code of Federal Regulations}
}
